\documentclass[conference]{IEEEtran}
\IEEEoverridecommandlockouts
\usepackage{lipsum} 
\usepackage[T1]{fontenc}
\usepackage{microtype}

\usepackage{graphicx}
\usepackage{cite}
\usepackage{amsfonts} 
\usepackage{amsmath}  
\usepackage{bm}
\usepackage{psfrag}
\usepackage{amsbsy}
\usepackage{amssymb}
\usepackage{mathrsfs}
\usepackage{stfloats}
\usepackage{dsfont}
\usepackage{setspace}
\usepackage{array}
\usepackage{bbm}
\usepackage{mathabx}
\usepackage{algorithmic}
\usepackage{algorithm}
\usepackage{glossaries}
\usepackage{color}
\usepackage{amsthm}
\usepackage{lettrine}

\newtheoremstyle{mystyle}
{}                
{}                
{\itshape}        
{}                
{\bfseries}       
{:}               
{.5em}            
{}                

\theoremstyle{mystyle}
\newtheorem{proposition}{Proposition}

\newtheorem{remark}{Remark}
\newcommand{\ds}{\displaystyle}

\usepackage{caption}
\DeclareMathOperator{\tr}{tr}

\def\BibTeX{{\rm B\kern-.05em{\sc i\kern-.025em b}\kern-.08em
		T\kern-.1667em\lower.7ex\hbox{E}\kern-.125emX}}
	
\usepackage[left=0.55in, right=0.55in, top=0.75in, bottom=1in]{geometry}

% ...

\usepackage{enumitem}
\setlist[itemize]{label=$\bullet$}

\usepackage{hyperref}

\hypersetup{
	colorlinks=true,       
	linkcolor=blue,       
	citecolor=green,      
	filecolor=magenta,    
	urlcolor=cyan         
}

\usepackage{caption}
\usepackage{url}
\makeatletter
\def\url@remove@extrahref#1{%
	\xdef\@savedhref{#1}%
	\expandafter\expandafter\expandafter\strip@url
	\expandafter\@savedhref
	\expandafter\@gobble
}
\makeatother

\makeglossaries

\newacronym{mac}{MAC}{multiple-access channel}
\newacronym{bc}{BC}{broadcast channel}
\newacronym{mimo}{MIMO}{multiple-input multiple-output}
\newacronym{siso}{SISO}{single-input single-output}
\newacronym{sc}{SC}{single-carrier}
\newacronym{mc}{MC}{multi-carrier}
\newacronym{ofdma}{OFDMA}{orthogonal frequency division multiple access}
\newacronym{af}{AF}{amplify-and-forward}
\newacronym{df}{DF}{decode-and-forward}
\newacronym{cf}{CF}{compress-and-forward}
\newacronym{mwrc}{MWRC}{multi-way relay channel}
\newacronym{pde}{PDE}{partial data exchange}
\newacronym{fde}{FDE}{full data exchange}
\newacronym{iid}{i.i.d.\@}{independent and identically distributed}
\newacronym{awgn}{AWGN}{additive white Gaussian noise}
\newacronym{awg}{AWG}{additive white Gaussian}
\newacronym{sic}{SIC}{successive interference cancellation}
\newacronym{snr}{SNR}{signal-to-noise ratio}
\newacronym{sinr}{SINR}{signal-to-interference-plus-noise ratio}
\newacronym{ber}{BER}{bit error rate}
\newacronym{zf}{ZF}{zero-forcing}
\newacronym{mmse}{MMSE}{minimum mean square error}
\newacronym{sud}{SUD}{single user decoding}
\newacronym{dof}{DoF}{degrees of freedom}
\newacronym{gdof}{GDoF}{generalized degrees of freedom}
\newacronym{nnc}{NNC}{noisy network coding}
\newacronym{dmn}{DMN}{discrete memoryless network}
\newacronym{csi}{CSI}{channel state information}
\newacronym{ee}{EE}{energy efficiency}
\newacronym{ian}{IAN}{treating interference as noise}
\newacronym{snd}{SND}{simultaneous non-unique decoding}
\newacronym{brd}{BRD}{best response dynamics}
\newacronym{br}{BR}{best response}
\newacronym{ne}{NE}{Nash equilibrium}
\newacronym{lhs}{LHS}{left-hand side}
\newacronym{rhs}{RHS}{right-hand side}
\newacronym{gee}{GEE}{global energy efficiency}
\newacronym{wsee}{WSEE}{weighted sum energy efficiency}
\newacronym{wpee}{WPEE}{weighted product energy efficiency}
\newacronym{wmee}{WMEE}{weighted minimum energy efficiency}
\newacronym{kkt}{KKT}{Karush-Kuhn-Tucker}
\newacronym{pc}{PC}{pseudo-concave}
\newacronym{qc}{QC}{quasi-concave}
\newacronym{ql}{QL}{quasi-linear}
\newacronym{pl}{PL}{pseudo-linear}
\newacronym{spc}{SPC}{strictly pseudo-concave}
\newacronym{sqc}{SQC}{strictly quasi-concave}
\newacronym{lfp}{LFP}{linear fractional problem}
\newacronym{clfp}{CLFP}{concave-linear fractional problem}
\newacronym{ccfp}{CCFP}{concave-convex fractional problem}
\newacronym{mmfp}{MMFP}{max-min fractional problem}
\newacronym{sorp}{SoRP}{sum-of-ratios problem}
\newacronym{porp}{PoRP}{product-of-ratios problem}
\newacronym{srp}{SRP}{single-ratio problem}
\newacronym{brb}{BRB}{branch-reduce-and-bound}
\newacronym{qos}{QoS}{quality-of-service}
\newacronym{comp}{CoMP}{cooperative multi-point}
\newacronym{ue}{UE}{user equipment}
\newacronym{bs}{BS}{base station}
\newacronym{mrc}{MRC}{maximum ratio combining}
\newacronym{d2d}{D2D}{device-to-device}
\newacronym{lmmse}{LMMSE}{linear minimum mean square error}
\newacronym{ris}{RIS}{reconfigurable intelligent surface}
\newacronym{svd}{SVD}{singular values decomposition}

\title{Secrecy Energy Efficiency Maximization in RIS-Aided Wireless Networks}

\author{
	Robert Kuku Fotock, {\em Student Member, IEEE}, Alessio Zappone, {\em Senior Member, IEEE},\\ Marco Di Renzo, {\em Fellow, IEEE}
	\thanks{R. K. Fotock and A. Zappone are with the University of Cassino and Southern Lazio, Cassino, Italy, and with CNIT, Parma, Italy (\{robertkuku.fotock,alessio.zappone\}@unicas.it). M. Di Renzo is with Universit\'e Paris-Saclay, CNRS, CentraleSup\'elec, Laboratoire des Signaux et Syst\`emes, 3 Rue Joliot-Curie, 91192 Gif-sur-Yvette, France. (marco.di-renzo@universite-paris-saclay.fr).  The work of R. K. Fotock was supported by the European Commission through the H2020-MSCA-ITN-METAWIRELESS project, grant agreement 956256. The work of A. Zappone was supported by the Project GARDEN, with CUP H53D23000480001 funded by EU in NextGenerationEU plan through the Italian "Bando Prin 2022 - D.D. 1409 del 14-09-2022" by MUR. The work of M. Di Renzo was supported in part by the EC HE projects COVER-101086228, UNITE-101129618, INSTINCT-101139161, and the ANR projects PEPR-NF-PERSEUS 22-PEFT-004 and PASSIONATE ANR-23-CHR4-0003-01.
}}

\begin{document}
	
	\sloppy  
	
	\maketitle
	
	\begin{abstract}
		
		This work proposes a provably convergent and low complexity optimization algorithm for the maximization of the secrecy energy efficiency in the uplink of a wireless network aided by a reconfigurable intelligent surface (RIS), in the presence of an eavesdropper. The mobile users' transmit powers and the RIS reflection coefficients are optimized. Numerical results show the performance of the proposed methods and compare the use of active and nearly-passive RISs from an energy-efficient perspective.

	\end{abstract}
	
	\vspace{1em}
	
	
	\section{Introduction}
\lettrine[nindent=0.1em,lines=2]{R}{ECONFIGURABLE} intelligent surfaces (RISs) have emerged as a major candidate technology for future wireless networks. In particular, they increase the rate with lower energy consumption than traditional antenna arrays, thanks to their simpler hardware \cite{DiRenzo2020, SmartWireless, RuiZhang_COMMAG, Huang2019}.  This potentially unlocks unprecedented energy efficiency (EE) gains, which is a major requirement in future wireless networks~\cite{Lopezperez2021survey}. On the other hand, the fact that RISs are not equipped with dedicated transmit amplifiers may limit their rate gains. Thus, active RISs have been put forth, which deploy analog amplifiers to increase the amplitude of the incoming signal~\cite{Long2021}. However, this leads to a fundamental trade-off in terms of EE, since the use of more complex hardware causes larger energy consumptions, too. In~\cite{Fotock2023}, the EE of active and nearly-passive RISs is compared, showing that active RISs do not always provide better EE.

Another key requirement in future wireless networks is security. Several contributions have appeared which study the use of RISs in conjunction with physical layer security. However, most available contributions focus on the maximization of the system secrecy rate, without addressing the system EE, i.e. the so-called secrecy energy efficiency (SEE)~\cite{ZapTSP13}. In the context of RIS-aided networks, available studies focus mainly on the system secrecy, without discussing EE aspects. A non-orthogonal multiple access (NOMA) network which employs a simultaneous transmission and reception (STAR) RIS is considered in~\cite{Li2023}, and the system secrecy outage probability is characterized. A NOMA-based network aided by a STAR-RIS is also considered in~\cite{Pei2023}, and
closed-form approximations of the secrecy outage probability are derived. Secrecy outage and average rate of an RIS-aided network are analyzed in~\cite{Trigui2021}, assuming that the RIS can apply discrete phase shifts. In~\cite{Xu2021}, analytical expressions of the ergodic secrecy rate of an RIS-aided network with multiple eavesdroppers are derived. In~\cite{Zhang2022}, a STAR-aided NOMA-based network is considered and the system worst-case secrecy capacity is maximized. In~\cite{Wei2022}, the ergodic secrecy capacity of an RIS-aided wireless network is analyzed and approximated in closed-form, considering flying eavesdroppers. In~\cite{Yang2020} the secrecy outage probability of an RIS-aided network is analyzed. In~\cite{Zhao2023}, the secrecy maximization rate of an RIS-aided network powered by wireless power transfer is optimized. In~\cite{Hoang2023} the secrecy rate of an RIS-aided network with space ground communications is optimized. In~\cite{Yizhi2023}, the sum secrecy rate of a multi-user RIS-aided wireless network is addressed. In~\cite{Cai2023}, secret communications aided by an RIS are implemented by a data interleaving method. 

While all the aforementioned works do not focus on the EE of secret communications, the SEE in RIS-aided networks is considered in a few works. In~\cite{Yichi2023},  a deep reinforcement learning method is employed to optimize the SEE of an RIS-aided network. In~\cite{Yang2023}, a combination of alternating maximization and sequential programming is employed to maximize the SEE of a multi-user network. A blend of sequential programming and alternating optimization is also used in \cite{Hao23}, wherein the minimum SEE of a multi-user network is optmized. While these few previous contributions on the SEE of RIS-aided networks focus on the use of nearly-passive RISs, this work introduces a new model that is general enough to encompass both active and nearly-passive RISs. Moreover, the proposed models is general enough to encompass a new and more general kind of RIS, namely RISs with global reflection capabilities. This new kind of RISs generalizes the use of traditional RISs with local reflection capabilities since the constraint on the reflected power is not applied to each reflecting element individually, but rather to the complete surface~\cite{Renzo2022}.

%
	
\section{System Model and Problem Formulation}\label{Sec:SysModel}

\begin{figure}[!t]
	\centering
	\includegraphics[width=0.46\textwidth]{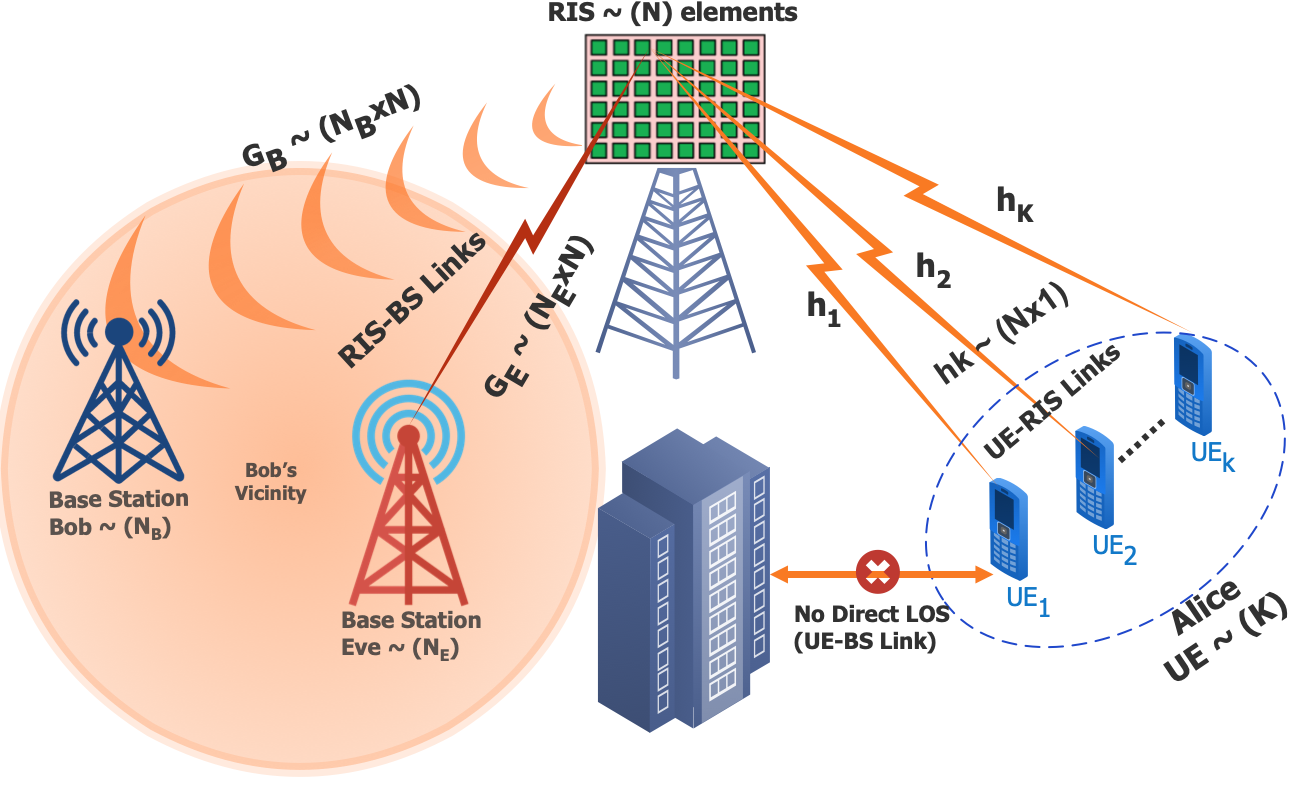}\caption{RIS-aided wireless network scenario} \label{Fig:scenario}
\end{figure}

Let us consider a network consisting of $K$ single-antenna mobile transmitters, labeled as Alices, which communicate with their receiver, labeled as Bob, equipped with $N_{B}$ antennas, through an RIS, equipped with $N$ reflecting elements. In the same area, we assume the presence of an eavesdropper, which is equipped with $N_{E}$ antennas and labelled as Eve, e.g. a neighboring base station from another operator.  
The considered scenario is depicted in Fig.~\ref{Fig:scenario}. Let us denote by $p_{k}$ the $k$-th user's transmit power, by $\gamma = \left(\boldsymbol{\gamma}_{1},\ldots,\boldsymbol{\gamma}_{N}\right)$ the $N \times 1$ vector containing the RIS refelction coefficients, by $\boldsymbol{h}_{k}$ the $N \times 1 $ channel between the $k$-th user and the RIS and by $\boldsymbol{G}_{B}$ and $\boldsymbol{G}_{E}$, the $N_{B} \times N$ and $N_{E} \times N$ channel from RIS to Bob and Eve, respectively. Given this setup, the SINR of user $k$ at the intended (Bob) and eavesdropping (Eve) receivers after applying the receive filters $\boldsymbol{c}_{k,B}$ and $\boldsymbol{c}_{k,E}$, respectively, are;

\begin{align}\label{Eq:SINR_B}
	\text{SINR}_{k,B}=\frac{p_{k}\left|\boldsymbol{c}_{k,B}^{H}\boldsymbol{A}_{k,B}\boldsymbol{\gamma}\right|^{2}}{\boldsymbol{c}_{k,B}^{H}\boldsymbol{W}_{B}\boldsymbol{c}_{k,B}+\sum_{m\neq k}p_{m}\left|\boldsymbol{c}_{k,B}^{H}\boldsymbol{A}_{m,B}\boldsymbol{\gamma}\right|^{2}}\;
\end{align}
\begin{align}\label{Eq:SINR_E}
	\text{SINR}_{k,E}=\frac{p_{k}\left|\boldsymbol{c}_{k,E}^{H}\boldsymbol{A}_{k,E}\boldsymbol{\gamma}\right|^{2}}{\boldsymbol{c}_{k,E}^{H}\boldsymbol{W}_{E}\boldsymbol{c}_{k,E}+\sum_{m\neq k}p_{m}\left|\boldsymbol{c}_{k,E}^{H}\boldsymbol{A}_{m,E}\boldsymbol{\gamma}\right|^{2}}\;
\end{align}
where $\boldsymbol{A}_{k,B}=\boldsymbol{G}_{B}\boldsymbol{H}_{k}$, $\boldsymbol{A}_{k,E}=\boldsymbol{G}_{E}\boldsymbol{H}_{k}$, while $\boldsymbol{W}_{B} = \sigma_{B}^{2}\boldsymbol{I}_{N_{B}} + \sigma_{RIS}^{2}\boldsymbol{G_{B}}\boldsymbol{\Gamma}\boldsymbol{\Gamma}^{H}\boldsymbol{G}_{B}^{H}$ and $\boldsymbol{W}_{E} = \sigma_{E}^{2}\boldsymbol{I}_{N_{E}} + \sigma_{RIS}^{2}\boldsymbol{G_{E}}\boldsymbol{\Gamma}\boldsymbol{\Gamma}^{H}\boldsymbol{G}_{E}^{H}$ are the overall covariance matrix of the intended and eavesdropping receivers, with $\boldsymbol{\Gamma} = \text{diag}\left(\boldsymbol{\gamma}\right)$, $\sigma^{2}_{\text{RIS}}$ the noise variance at the RIS, while $\sigma^{2}_{\text{B}}$ and $\sigma^{2}_{\text{E}}$ the noise variances at the legitimate and eavesdropping receivers, respectively.
Thus, the system secrecy rate is given by
\begin{align}\label{Eq:SecRate}
	&R_{S} \!=\! \left[\sum_{k}\left(R_{k,B}-R_{k,E}\right)\right]^+ \!\!\!=\! \max\left\{\sum_{k}\left(R_{k,B}-R_{k,E}\right),0\!\right\} \notag \\
	&\!=\! \max\!\left\{\sum_{k=1}^{K}\log_{2}\left(1\!+\!\text{SINR}_{k,B}\right)\! -\! \log_{2}\left(1\!+\!\text{SINR}_{k,E}\right),0\!\right\}
\end{align}
As for the power consumption of the legitimate system, the radio frequency power consumed by the RIS is given by the difference between the incident power on the RIS $P_{in}$ and the power that departs from the RIS, $P_{out}$, which after some elaborations~\cite{Robert2023}, can be computed as
\begin{align}\label{Eqn:P_RF}
	P_{out} - P_{in} &= \tr\left(\sum_{k=1}^{K}p_{k}\boldsymbol{\Gamma}\boldsymbol{h}_{k}\boldsymbol{h}^{H}_{k}\boldsymbol{\Gamma}^{H} +  \sigma^{2}_{\text{RIS}}\boldsymbol{\Gamma}\boldsymbol{\Gamma}^{H}\right)\\
	&- \sum_{k=1}^{K}p_{k} \left\Vert \boldsymbol{h}_{k}\right\Vert^{2} - \sigma^{2}_{\text{RIS}}N = \tr\left(\left(\boldsymbol{\gamma}\boldsymbol{\gamma}^{H}-\boldsymbol{I}_{N}\right)\boldsymbol{R}\right)\notag
\end{align}

\noindent wherein $\boldsymbol{R}=\textstyle\sum_{k=1}^{K} p_{k}\boldsymbol{H}_{k}^{H}\boldsymbol{H}_{k}+\sigma_{\text{RIS}}^{2}\boldsymbol{I}_{N}$.  In addition, the total power consumption $P_{tot}$ is obtained by summing the radio-frequency power consumed by the RIS and the users’ transmit powers, as well as the static power consumption of the whole legitimate system, which yields $P_{tot} = P_{out} - P_{in} + \sum_{k=1}^{K}\mu_{k}p_{k} + P_{c}$ wherein $\mu_{k}$ symbolizes the inverse efficiency of the transmit amplifier associated with the $k$-th transmitter and $P_{c} = NP_{c,n} + P^{\text{RIS}}_{0} + P_{0}$, with $P_{c,n} $ the static power consumption of the $n$-th RIS element, $P^{\text{RIS}}_{0}$ is the rest of the static power consumed by the RIS and $P_{0}$ encompasses all other sources of power consumption in the legitimate system. The considered power consumption model has been developed for an active RIS, but it is general enough to be applied to the case of a nearly-passive RIS as a special case. Specifically, in the case of a nearly-passive RIS, $P_{out} \leq P_{in}$ and thus the difference $P_{out} - P_{in}$ does not appear in the power consumption $P_{tot}$. Moreover, if a nearly-passive RIS is employed, the terms $P_{c,n}$ and $ P^{\text{RIS}}_{0}$ will have a lower numerical value than in the case of an active RIS, due to the fact that simpler hardware is employed in nearly-passive RIS, i.e. no analog amplifiers are used.

\begin{remark}\label{remark:constraint_state}
	The constraints that should be enforced on the RIS vector $\boldsymbol{\gamma}$ depend on whether the RIS is active or nearly-passive. Specifically, if the RIS is active, then it must hold $0 \leq P_{out} - P_{in} \leq P_{R,\text{max}}$ with $P_{R,\text{max}}$ the maximum radio-frequency power that the RIS amplifier can provide, as shown in~\eqref{Prob:bSEE_ARIS}. If instead, the RIS is nearly-passive, the constraint reduces to the more traditional expression $P_{out} \leq P_{in}$. It can be seen that the constraint in the nearly-passive case can be obtained as a special case of the constraint~\eqref{Prob:bSEE_ARIS} in the active case, by relaxing the
	first inequality and setting $\sigma^{2}_{\text{RIS}} = 0\, \text{and} \, P_{R,\text{max}} = 0$ in the second inequality. Thus, in the following we will focus on the more general active RIS scenario, keeping in mind that the optimization techniques
	that will be developed are able to tackle the SEE maximization problem with a nearly-passive RIS, too.
\end{remark}
Finally, the SEE is given by the ratio between the secrecy rate and total power consumption, namely $\text{SEE} = R_{S}/P_{tot}$ and the SEE maximization problem is cast as
\begin{subequations}\label{Prob:SEE_ARIS}
	\begin{align}
		&\ds\max_{\boldsymbol{\gamma},\boldsymbol{p},\boldsymbol{C}}\; \text{SEE}(\boldsymbol{\gamma},\boldsymbol{p},\boldsymbol{C})\label{Prob:aSEE_ARIS}\\
		&\;\text{s.t.}\;\tr\left(\boldsymbol{R}\right)\leq \tr(\boldsymbol{R}\boldsymbol{\gamma}\boldsymbol{\gamma}^{H})\leq P_{R,max}+\tr\left(\boldsymbol{R}\right)\label{Prob:bSEE_ARIS}\\
		&\;\quad\;\;0\leq p_{k}\leq P_{max,k}\;\forall\;k=1,\ldots,K\;,\label{Prob:cSEE_ARIS}
	\end{align}
\end{subequations}
with $\boldsymbol{C} = \left[\boldsymbol{c}_{1},\ldots,\boldsymbol{c}_{K} \right]$. We observe that Problem~\eqref{Prob:SEE_ARIS} is always feasible, since setting $\left|\gamma_{n}\right|=1$ for all $n$ fulfills all constraints.

\section{SEE Maximization}\label{Sec:SEE_Max}
In this section, we tackle the SEE maximization problem in~\eqref{Prob:SEE_ARIS}. Our approach  hinges on the alternating optimization of the RIS reflection vector $\boldsymbol{\gamma}$, the users' transmit powers $\boldsymbol{p}$, and the linear receive filters $\boldsymbol{C}$. These three subproblems are treated separately in the next three subsections. Afterwards, the overall resource allocation algorithm is presented, and its convergence and complexity is analyzed. 

\subsection[Optimization of gamma]{Optimization of $\boldsymbol{\gamma}$}
Considering the RIS vector $\boldsymbol{\gamma}$, the problem is expressed as:
\begin{subequations}\label{Prob:SEC_ARIS_gamma}
	\begin{align}
		&\ds\max_{\boldsymbol{\gamma}}\; B\frac{\textstyle\sum_{k=1}^{K}\log_{2}\left(1+\text{SINR}_{k,B}\right) - \log_{2}\left(1+\text{SINR}_{k,E}\right)}{\tr\left(\boldsymbol{R}\boldsymbol{\gamma}\boldsymbol{\gamma}^{H}\right)+P_{c,eq}} \label{Prob:aSEC_ARIS_gamma}\\
		&\;\text{s.t.}\;\tr\left(\boldsymbol{R}\right)\leq \tr(\boldsymbol{R}\boldsymbol{\gamma}\boldsymbol{\gamma}^{H})\leq P_{R,max}+\tr\left(\boldsymbol{R}\right)\label{Prob:bSEC_ARIS_gamma}
	\end{align}
\end{subequations}
wherein $P_{c,eq}=\sum_{k}p_{k}\mu_{k}+P_{c}-\tr(\boldsymbol{R})$. Even if formulated only with respect to the vector $\boldsymbol{\gamma}$, the problem is quite challenging since the objective is non-concave and not even pseudo-concave, and the first inequality in~\eqref{Prob:bSEC_ARIS_gamma}  is a non-convex constraint. Moreover, direct application of fractional programming techniques is not feasible, since the numerator of~\eqref{Prob:aSEC_ARIS_gamma} is not inherently concave~\cite{ZapNow15}. To circumvent this challenge, we employ the sequential fractional programming method. To this end, we commence by
expressing the term $\boldsymbol{c}_{k,i}^{H}\boldsymbol{W}_{i}\boldsymbol{c}_{k,i}$ in terms of the vector $\boldsymbol{\gamma}$, instead of matrix $\boldsymbol{\Gamma}$. To achieve this, we define $\boldsymbol{u}_{k,i}=\boldsymbol{G}_{i}^{H}\boldsymbol{c}_{k,i}$ and $\widetilde{\boldsymbol{U}}_{k,i}=\text{diag}\left(|u_{1,i}|^{2},\ldots,|u_{N,i}|^{2}\right)$. Subsequently, by incorporating the expression of $\boldsymbol{W}_{i}$\footnote{Note that, for simplicity, we have considered the subscript $i$  to represent the corresponding receiver, i.e. Bob or Eve.}, we obtain $\boldsymbol{c}_{k,i}^{H}\boldsymbol{W}_{i}\boldsymbol{c}_{k,i}=\sigma^{2}\|\boldsymbol{c}_{k,i}\|^{2}+\sigma_{\text{RIS}}^{2}\boldsymbol{\gamma}^{H}\widetilde{\boldsymbol{U}}_{k,i}\boldsymbol{\gamma}$. Then, a concave lower-bound of the numerator in \eqref{Prob:aSEC_ARIS_gamma} is required. We utilize the bound~\cite{Mohamed2023}
\begin{align}
	\log\left(1\!+\!\frac{x}{y}\right)\!\geq\! \log\left(1\!+\!\frac{\bar{x}}{\bar{y}}\right)\!+\!\frac{\bar{x}}{\bar{y}}\left(\frac{2\sqrt{x}}{\sqrt{\bar{x}}}\!-\!\frac{x+y}{\bar{x}\!+\!\bar{y}}\!-\!1\right)
\end{align}
which holds for any $x$, $y$, $\bar{x}$ and $\bar{y}$, and holds with equality whenever $x=\bar{x}$ and $y=\bar{y}$. Using the feasible vector $\bar{\boldsymbol{\gamma}}$ for the RIS reflection coefficients and applying the bound to the data rate of each user, we get for the $k$th user

\begin{align}\label{Eq:BarR_Active}
	\bar{R}_{k,i}&=\bar{A}_{k,i}+\bar{B}_{k,i}\Bigg(\bar{D}_{k,i}|\boldsymbol{c}_{k,i}^{H}\boldsymbol{A}_{k,i}\boldsymbol{\gamma}|-\bar{F}_{k,i}\notag\\
	&-\bar{E}_{k,i}\left(\sigma_{\text{RIS}}^{2}\boldsymbol{\gamma}^{H}\widetilde{\boldsymbol{U}}_{k,i}\boldsymbol{\gamma}+\textstyle\sum_{m=1}^{K}p_{m}|\boldsymbol{c}_{k,i}^{H}\boldsymbol{A}_{m,i}\boldsymbol{\gamma}|^{2}\right)\Bigg)
\end{align}
wherein
\begin{align}
	\bar{A}_{k,i}\! &= \!\log_{2}\!\!\left(\!\!1\!+\!\frac{\!p_{k}|\boldsymbol{c}_{k,i}^{H}\boldsymbol{A}_{k,i}\bar{\boldsymbol{\gamma}}|^{2}}{\sigma^{2}\|\boldsymbol{c}_{k,i}\|^{2}\!+\!\sigma_{\text{RIS}}^{2}\bar{\boldsymbol{\gamma}}^{H}\widetilde{\boldsymbol{U}}_{k,i}\bar{\boldsymbol{\gamma}}\!+\!\!\!\!\displaystyle{\sum_{m\neq k}p_{m}|\boldsymbol{c}_{k,i}^{H}\boldsymbol{A}_{m,i}\bar{\boldsymbol{\gamma}}|^{2}}}\!\!\right)\notag
	\end{align}
	\begin{align}
	\bar{B}_{k,i}&=\frac{p_{k}|\boldsymbol{c}_{k,i}^{H}\boldsymbol{A}_{k,i}\bar{\boldsymbol{\gamma}}|^{2}}{\sigma^{2}\|\boldsymbol{c}_{k,i}\|^{2}+\sigma_{\text{RIS}}^{2}\bar{\boldsymbol{\gamma}}^{H}\widetilde{\boldsymbol{U}}_{k,i}\bar{\boldsymbol{\gamma}}+\sum_{m\neq k}p_{m}|\boldsymbol{c}_{k,i}^{H}\boldsymbol{A}_{m,i}\bar{\boldsymbol{\gamma}}|^{2}}\notag\\
	\bar{D}_{k,i}&=2/|\boldsymbol{c}_{k,i}^{H}\boldsymbol{A}_{k,i}\bar{\boldsymbol{\gamma}}|\;,\bar{F}_{k,i}=\bar{E}_{k,i}\sigma^{2}\|\boldsymbol{c}_{k,i}\|^{2}+1\notag\\
	\bar{E}_{k,i}&=\frac{1}{\sigma^{2}\|\boldsymbol{c}_{k,i}\|^{2}+\sigma_{\text{RIS}}^{2}\bar{\boldsymbol{\gamma}}^{H}\widetilde{\boldsymbol{U}}_{k,i}\bar{\boldsymbol{\gamma}}+\sum_{m=1}^{K}p_{m}|\boldsymbol{c}_{k,i}^{H}\boldsymbol{A}_{m,i}\bar{\boldsymbol{\gamma}}|^{2}}\notag
\end{align}
\vspace{-0.05cm}
Each term of the numerator in \eqref{Prob:SEC_ARIS_gamma} represents the secrecy rate for user $k$. The lower-bound approximation for user $k$ is:
\begin{align}\label{Eq:BarSecR_Active}
	\widebar{SR}_{k} &= \bar{R}_{k,i=\text{Bob}} -  \bar{R}_{k,i=\text{Eve}}\\
	&= \bar{A}_{k,B}-\bar{A}_{k,E}+\bar{B}_{k,B}\bar{D}_{k,B}|\boldsymbol{c}_{k,B}^{H}\boldsymbol{A}_{k,B}\boldsymbol{\gamma}|\notag\\
	&- \bar{B}_{k,E}\bar{D}_{k,E}|\boldsymbol{c}_{k,E}^{H}\boldsymbol{A}_{k,E}\boldsymbol{\gamma}| - \bar{B}_{k,B}\bar{F}_{k,B} + \bar{B}_{k,E}\bar{F}_{k,E}\notag\\
	&-\bar{B}_{k,B}\bar{E}_{k,B}\left(\sigma_{\text{RIS}}^{2}\boldsymbol{\gamma}^{H}\widetilde{\boldsymbol{U}}_{k,B}\boldsymbol{\gamma}+\textstyle\sum_{m=1}^{K}p_{m}|\boldsymbol{c}_{k,B}^{H}\boldsymbol{A}_{m,B}\boldsymbol{\gamma}|^{2}\right)\notag\\
	&+\bar{B}_{k,E}\bar{E}_{k,E}\left(\sigma_{\text{RIS}}^{2}\boldsymbol{\gamma}^{H}\widetilde{\boldsymbol{U}}_{k,E}\boldsymbol{\gamma}+\textstyle\sum_{m=1}^{K}p_{m}|\boldsymbol{c}_{k,E}^{H}\boldsymbol{A}_{m,E}\boldsymbol{\gamma}|^{2}\right)\notag
\end{align}
Notably, the terms $-|\boldsymbol{c}_{k,E}^{H}\boldsymbol{A}_{k,E}\boldsymbol{\gamma}|$, $-\sigma_{\text{RIS}}^{2}\boldsymbol{\gamma}^{H}\widetilde{\boldsymbol{U}}_{k,B}\boldsymbol{\gamma}$ and $-\textstyle\sum_{m=1}^{K}p_{m}|\boldsymbol{c}_{k,B}^{H}\boldsymbol{A}_{m,B}\boldsymbol{\gamma}|^{2}$ are concave and thus the only non-concave terms in $\widebar{SR}_{k}$ are $|\boldsymbol{c}_{k,B}^{H}\boldsymbol{A}_{k,B}\boldsymbol{\gamma}|$, $\sigma_{\text{RIS}}^{2}\boldsymbol{\gamma}^{H}\widetilde{\boldsymbol{U}}_{k,E}\boldsymbol{\gamma}$ and $\textstyle\sum_{m=1}^{K}p_{m}|\boldsymbol{c}_{k,E}^{H}\boldsymbol{A}_{m,E}\boldsymbol{\gamma}|^{2}$. However, since these terms are convex in $\boldsymbol{\gamma}$, their first-order Taylor expansion around any point $\bar{\boldsymbol{\gamma}}$ provides a concave lower-bound as follows:
\begin{align}\label{Eq:BarApproxSecR_Active}
SR_{k}&\geq \widebar{SR}_{k}\geq \bar{A}_{k,B}-\bar{A}_{k,E}+\bar{B}_{k,B}\bar{D}_{k,B}\mathscr{T}\left\{Q_{k,B}\left(\boldsymbol{\gamma}\right)\right\}\notag\\
&- \bar{B}_{k,E}\bar{D}_{k,E}Q_{k,E}\left(\boldsymbol{\gamma}\right) - \bar{B}_{k,B}\bar{F}_{k,B} + \bar{B}_{k,E}\bar{F}_{k,E}\notag\\
&-\bar{B}_{k,B}\bar{E}_{k,B}\left(V^{(1)}_{k,B} \left(\boldsymbol{\gamma}\right)+ V^{(2)}_{k,B}\left(\boldsymbol{\gamma}\right)\right)\notag\\
&+\bar{B}_{k,E}\bar{E}_{k,E}\mathscr{T}\left\{V^{(1)}_{k,E} \left(\boldsymbol{\gamma}\right)+ V^{(2)}_{k,E}\left(\boldsymbol{\gamma}\right)\right\}\!=\!\widetilde{SR}_{k}
\end{align}  
wherein, we have defined the following terms and their first-order Taylor expansion around any feasible point $\bar{\boldsymbol{\gamma}}$ as
\begin{align*}
	& Q_{k,i}\left(\boldsymbol{\gamma}\right) = |\boldsymbol{c}_{k,i}^{H}\boldsymbol{A}_{k,i}\boldsymbol{\gamma}|\notag\\
	&\mathscr{T}\left\{Q_{k,i}\left(\boldsymbol{\gamma}\right)\right\} =|\boldsymbol{c}_{k,i}^{H}\boldsymbol{A}_{k,i}\bar{\boldsymbol{\gamma}}| +\Re\left\{\frac{\boldsymbol{A}_{k,i}^{H}\boldsymbol{c}_{k,i}\boldsymbol{c}_{k,i}^{H}\boldsymbol{A}_{k,i}\bar{\boldsymbol{\gamma}}}{|\boldsymbol{c}_{k,i}^{H}\boldsymbol{A}_{k,i}\bar{\boldsymbol{\gamma}}|}(\boldsymbol{\gamma}-\bar{\boldsymbol{\gamma}})\right\}\notag\\
	&V^{(1)}_{k,i}\left(\boldsymbol{\gamma}\right) = \sigma_{\text{RIS}}^{2}\boldsymbol{\gamma}^{H}\widetilde{\boldsymbol{U}}_{k,i}\boldsymbol{\gamma},\;V^{(2)}_{k,i}\left(\boldsymbol{\gamma}\right) = \textstyle\sum_{m=1}^{K}p_{m}|\boldsymbol{c}_{k,i}^{H}\boldsymbol{A}_{m,i}\boldsymbol{\gamma}|^{2}\quad \notag\\
	&\mathscr{T}\left\{V^{(1)}_{k,i}\left(\boldsymbol{\gamma}\right)\right\} = \sigma_{\text{RIS}}^{2}\bar{\boldsymbol{\gamma}}^{H}\widetilde{\boldsymbol{U}}_{k,i}\bar{\boldsymbol{\gamma}} + 2\sigma_{RIS}^{2}\Re\left\{ \widetilde{\boldsymbol{U}}_{k,i}\bar{\boldsymbol{\gamma}}(\boldsymbol{\gamma}-\bar{\boldsymbol{\gamma}})\right\}\notag\\
	&\mathscr{T}\left\{V^{(2)}_{k,i}\left(\boldsymbol{\gamma}\right)\right\} = \textstyle\sum_{m=1}^{K}p_{m}|\boldsymbol{c}_{k,i}^{H}\boldsymbol{A}_{m,i}\bar{\boldsymbol{\gamma}}|^{2}\notag\\
	 &\;\qquad \qquad \qquad \; + 2\textstyle\sum_{m=1}^{K}p_{m}\Re\left\{ \boldsymbol{A}_{m,i}^{H}\boldsymbol{c}_{k,i}\boldsymbol{c}_{k,i}^{H}\boldsymbol{A}_{m,i}\bar{\boldsymbol{\gamma}}(\boldsymbol{\gamma}-\bar{\boldsymbol{\gamma}})\right\}\notag
\end{align*}
\vspace{-0.05cm}
Lastly, a lower-bound for \eqref{Prob:bSEC_ARIS_gamma} is obtained by noticing that $\tr\left(\boldsymbol{R}\boldsymbol{\gamma}\boldsymbol{\gamma}^{H}\right)$ is convex in $\boldsymbol{\gamma}$. Then, its first-order Taylor expansion around any point $\bar{\boldsymbol{\gamma}}$ provides a lower-bound, as follows:
\begin{align}
	\tr\left(\boldsymbol{R}\boldsymbol{\gamma}\boldsymbol{\gamma}^{H}\right) &= \boldsymbol{\gamma}\boldsymbol{R}\boldsymbol{\gamma}^{H} \geq \bar{\boldsymbol{\gamma}}\boldsymbol{R}\bar{\boldsymbol{\gamma}}^{H} + 2\Re\left\{\bar{\boldsymbol{\gamma}}^{H}\boldsymbol{R}\left(\boldsymbol{\gamma} - \bar{\boldsymbol{\gamma}}\right)\right\}
\end{align}
Consequently, each iteration of the sequential method solves:
\begin{subequations}\label{Prob:SEE_Gamma_Active}
	\begin{align}
		&\ds\max_{\boldsymbol{\gamma}}\;\frac{\textstyle\sum_{k=1}^{K}\widetilde{SR}_{k}}{\tr\left(\boldsymbol{R}\boldsymbol{\gamma}\boldsymbol{\gamma}^{H}\right)+P^{\text{(a)}}_{c,eq}} \label{Prob:aSEE_Gamma_Active}\\
		&\;\text{s.t.}\;\boldsymbol{\gamma}\boldsymbol{R}\boldsymbol{\gamma}^{H} \leq P_{R,max}+\tr\left(\boldsymbol{R}\right)\label{Prob:bSEE_Gamma_Active}\\
		&\; \quad \;\bar{\boldsymbol{\gamma}}\boldsymbol{R}\bar{\boldsymbol{\gamma}}^{H} + 2\Re\left\{\bar{\boldsymbol{\gamma}}^{H}\boldsymbol{R}\left(\boldsymbol{\gamma} - \bar{\boldsymbol{\gamma}}\right)\right\} \geq \tr\left(\boldsymbol{R}\right)\label{Prob:cSEE_Gamma_Active}
	\end{align}
\end{subequations}
The objective in \eqref{Prob:SEC_ARIS_gamma} has a concave numerator and a convex denominator. Thus, \eqref{Prob:SEC_ARIS_gamma} is a pseudo-concave maximization with convex constraints, which can be solved by fractional programming. The procedure is stated in Algorithm~\eqref{Alg:SCA_gamma1}.
\begin{algorithm}
	\caption{RIS optimization}
	\begin{algorithmic}\label{Alg:SCA_gamma1}
		\STATE $\epsilon > 0$, $\bar{\boldsymbol{\gamma}}=\boldsymbol{\gamma}_{0}$ \texttt{with} $\boldsymbol{\gamma}_{0}$ \texttt{any feasible vector};
		\REPEAT
		\STATE  $\bar{\boldsymbol{\gamma}}=\boldsymbol{\gamma}_{0}$, with $\boldsymbol{\gamma}_{0}$ \texttt{the solution of} \eqref{Prob:SEE_Gamma_Active}; 
		\UNTIL{$\|\bar{\boldsymbol{\gamma}}-\boldsymbol{\gamma}_{0}\|<\epsilon$}
	\end{algorithmic}
\end{algorithm}

\begin{proposition}\label{Prop:SCA_gamma1}
	Algorithm \ref{Alg:SCA_gamma1} monotonically improves the value of the objective and converges to a point fulfilling the Karush-Kuhn-Tucker (KKT) optimality conditions in \eqref{Prob:SEC_ARIS_gamma}.
\end{proposition}
\begin{IEEEproof}
	The proof follows noticing that Algorithm~\eqref{Alg:SCA_gamma1} fulfills all assumptions of sequential fractional programming.
\end{IEEEproof}

\vspace{-0.37cm}
\subsection[Optimization of transmit powers]{Optimization of $\boldsymbol{p}$}
Let us define $a^{(i)}_{k,m} = |\boldsymbol{c}_{k,i}^{H}\boldsymbol{A}_{m,i}\boldsymbol{\gamma}|^{2}$ for all $m$ and $k$, $d_{k,i} = \boldsymbol{c}_{k,i}^{H}\boldsymbol{W}_{i}\boldsymbol{c}_{k,i}$, $P_{c,eq} = \sigma_{\text{RIS}}^{2}\left(\|\boldsymbol{\gamma}\|^{2}-N\right) + P_{c}$ and $\mu_{k,eq} = \mu_{k} - \|\boldsymbol{h}_{k}\|^{2} + |\boldsymbol{H}_{k}\boldsymbol{\gamma}|^{2}$. With these definitions, the power optimization problem can be formulated as:

\begin{subequations}\label{Prob:SEE_ARIS_power}
	\begin{align}
		&\ds\max_{\boldsymbol{p}}\,\frac{\sum_{k=1}^{K}\ds\log_{2}\left(1+\frac{p_{k}a^{(B)}_{k,k}}{d_{k,B}+\sum_{m\neq k}p_{m}a^{(B)}_{k,m}}\right)}{\sum_{k=1}^{K}\mu_{k,eq}p_{k}+P_{c,eq}}\notag\\
		&-\frac{\sum_{k=1}^{K}\ds\log_{2}\left(1+\frac{p_{k}a^{(E)}_{k,k}}{d_{k,E}+\sum_{m\neq k}p_{m}a^{(E)}_{k,m}}\right)}{\sum_{k=1}^{K}\mu_{k,eq}p_{k}+P_{c,eq}}\label{Prob:aSEE_ARIS_power}\\
		&\;\text{s.t.}\;0\leq p_{k}\leq P_{max,k}\;,\forall\; k=1,\ldots,K.
	\end{align}
\end{subequations}

However, the objective in \eqref{Prob:aSEE_ARIS_power} is not pseudo-concave due to the non-concavity of its numerator. As a result, the application of conventional fractional programming methods, such as the one discussed in \cite{ZapNow15}, are not computationally feasible for solving \eqref{Prob:aSEE_ARIS_power}. To overcome this challenge, we employ the sequential fractional programming method, also proposed in~\cite{ZapNow15}. This approach provides a pseudo-concave lower-bound for \eqref{Prob:aSEE_ARIS_power}, enabling us to maximize it using fractional programming. The detailed formulation is as follows:

\begin{align}\label{Prob:SEE_ARIS_power_2}
	\text{SEE}(\boldsymbol{p}) &= g_{1, B}(\boldsymbol{p}) - g_{2, B}(\boldsymbol{p}) - g_{1, E}(\boldsymbol{p}) + g_{2, E}(\boldsymbol{p})\notag\\	
	&= g_{1,B}(\boldsymbol{p})+ g_{2,E}(\boldsymbol{p}) -  g_{2, B}(\boldsymbol{p}) - g_{1, E }(\boldsymbol{p}) 
\end{align}
wherein we denote

\begin{align}\label{Prob:SEE_ARIS_power_funcs}
	&g_{1, i}(\boldsymbol{p})  = \frac{\sum_{k=1}^{K}\log_{2}\left(\!d_{k,i}+\sum_{k=1}^{K}p_{k,i}a^{(i)}_{k,k}\right)}{\sum_{k=1}^{K}\mu_{k,eq}p_{k}+P_{c,eq}}\notag\\ 
	&g_{2, i}(\boldsymbol{p})  = \frac{\sum_{k=1}^{K}\log_{2}\left(d_{k,i}+\sum_{m\neq k}p_{m}a^{(i)}_{k,m}\right)}{\sum_{k=1}^{K}\mu_{k,eq}p_{k}+P_{c,eq}}\notag\;
\end{align}
where $i$ denotes the specific receiver (either Bob or Eve). Then, consider $f_{1,i}(\boldsymbol{p})$ and $f_{2,i}(\boldsymbol{p})$ as the respective numerators of $g_{1,i}(\boldsymbol{p})$ and $g_{2,i}(\boldsymbol{p})$. Notably, while $f_{1,B}(\boldsymbol{p})$ and $f_{2,E}(\boldsymbol{p})$ are concave in $\boldsymbol{p}$, $-f_{2,B}(\boldsymbol{p})$ and $-f_{1,E}(\boldsymbol{p})$ render the numerator of the SEE non-concave.
Nevertheless, we can derive a pseudo-concave lower-bound for $\text{SEE}(\boldsymbol{p})$, denoted as $\widetilde{\text{SEE}}(\boldsymbol{p})$, by replacing $f_{2, B}(\boldsymbol{p})$ and $f_{1, E}(\boldsymbol{p})$ with their first-order Taylor expansion around any feasible point $\bar{\boldsymbol{p}}$. Thus,~\eqref{Prob:SEE_ARIS_power_2} becomes
\begin{align}
	&\text{SEE}(\boldsymbol{p}) \geq g_{1,B}(\boldsymbol{p})+ g_{2,E}(\boldsymbol{p}) -  g_{2, B}(\bar{\boldsymbol{p}}) - g_{1, E}(\bar{\boldsymbol{p}})\notag\\ 
	&-  \frac{\left(\nabla f_{2,B}(\boldsymbol{p})\right)^{T}\left(\boldsymbol{p}-\bar{\boldsymbol{p}}\right)}{\sum_{k=1}^{K}\mu_{k,eq}p_{k}+P_{c,eq}} - \frac{\left(\nabla f_{1,E}(\boldsymbol{p})\right)^{T}\left(\boldsymbol{p}-\bar{\boldsymbol{p}}\right)}{\sum_{k=1}^{K}\mu_{k,eq}p_{k}+P_{c,eq}} = \widetilde{\text{SEE}}(\boldsymbol{p})
\end{align}
wherein for all $j=1,\ldots,K$, it holds
\begin{align}
	\frac{\partial f_{2,B}}{\partial p_{j}}&=\sum_{k\neq j}\frac{a^{(B)}_{k,j}}{d_{k,B}+\sum_{m\neq k}p_{m}a^{(B)}_{k,m}}\\
	\frac{\partial f_{1,E}}{\partial p_{j}}&=\sum_{k = 1}^{K}\frac{a^{(B)}_{k,j}}{d_{k,B}+\sum_{m=1}^{K}p_{m}a^{(B)}_{k,m}}
\end{align}
A sequential fractional programming algorithm can be devised to address \eqref{Prob:SEE_ARIS_power}, in which each iteration solves the problem
\begin{subequations}\label{Prob:SEE_ARIS_power_approx}
	\begin{align}
		&\ds\max_{\boldsymbol{p}}\,\widetilde{\text{SEE}}(\boldsymbol{p})\\
		&\;\text{s.t.}\;0\leq p_{k}\leq P_{max,k}\;,\forall\; k=1,\ldots,K.
	\end{align}
\end{subequations}
To address Problem~\eqref{Prob:SEE_ARIS_power_approx}, which involves maximizing a pseudo-concave function, one can employ fractional programming techniques \cite{ZapNow15}.

\begin{algorithm}
	\caption{Power optimization}
	\begin{algorithmic}\label{Alg:SCA_p1}
		\STATE $\epsilon > 0$, $\bar{\boldsymbol{p}}=\boldsymbol{p}_{0}$ \texttt{with} $\boldsymbol{p}_{0}$ \texttt{any feasible vector};
		\REPEAT
		\STATE $\bar{\boldsymbol{p}}=\boldsymbol{p}_{0}$, with $\boldsymbol{p}_{0}$ \texttt{the solution of} \eqref{Prob:SEE_ARIS_power_approx};
		\UNTIL{$\|\bar{\boldsymbol{p}}-\boldsymbol{p}_{0}\|<\epsilon$}
	\end{algorithmic}
\end{algorithm}
\begin{proposition}\label{Prop:SCA_p1}
	Algorithm \ref{Alg:SCA_p1} monotonically improves the value of the objective and converges to a point fulfilling the KKT optimality conditions of \eqref{Prob:SEE_ARIS_power_approx}.
\end{proposition}
\begin{IEEEproof}
	The proof follows as for Proposition \ref{Prop:SCA_gamma1}. 
\end{IEEEproof}

\subsection[Optimization of the LMMSE filters]{Optimization of $\boldsymbol{C}$}
The optimization of the receive filters in $\boldsymbol{C}$, influences exclusively the numerator of the SEE. Furthermore, it can be independently decoupled across the users. The well-established solution to this problem is the linear minimum mean squared error (MMSE) receiver, which, for the considered case, is given by $\boldsymbol{c}_{k,i}=\sqrt{p}_{k}\boldsymbol{M}_{k,i}^{-1}\boldsymbol{A}_{k,i}\boldsymbol{\gamma}$ where $\boldsymbol{M}_{k,i}=\sum_{m\neq k}p_{m}\boldsymbol{A}_{m,i}\boldsymbol{\gamma}\boldsymbol{\gamma}^{H}\boldsymbol{A}_{m,i}^{H}+\boldsymbol{W}_{i}$ represents the interference-plus-noise covariance matrix of user $k$.

\subsection[Overall Algorithm, Convergence and Complexity]{Overall  Algorithm, Convergence, and  Complexity}
The overall alternating maximization algorithm can be stated as in Algorithm~\ref{Alg:SEE1}, and the following result holds.
\begin{algorithm}[!h]
	\caption{Solution algorithm for Problem \eqref{Prob:SEE_ARIS}}
	\begin{algorithmic}\label{Alg:SEE1}
		\STATE \texttt{Set} $\epsilon > 0$, \texttt{initialize} $\tilde{\boldsymbol{p}}$, $\tilde{\boldsymbol{\gamma}}$ \texttt{to 
			feasible values}
		\STATE $\boldsymbol{c}_{k}=\sqrt{p}_{k}\boldsymbol{M}_{k}^{-1}\boldsymbol{A}_{k}$ \texttt{for all} $k$;
		\REPEAT
		\STATE $\text{SEE}_{in}=\text{SEE}(\tilde{\boldsymbol{p}},\tilde{\boldsymbol{\gamma}},\boldsymbol{C})$; \texttt{Given} $\tilde{\boldsymbol{p}}$ \texttt{run} Algorithm \ref{Alg:SCA_gamma1};
		\STATE \texttt{Let} $\tilde{\boldsymbol{\gamma}}$ \texttt{be the optimized RIS vector};
		\STATE \texttt{Given} $\tilde{\boldsymbol{\gamma}}$ \texttt{run} Algorithm \ref{Alg:SCA_p1};
		\STATE \texttt{Let} $\tilde{\boldsymbol{p}}$ \texttt{be the optimized power vector};
		\STATE $\boldsymbol{c}_{k}=\sqrt{p}_{k}\boldsymbol{M}_{k}^{-1}\boldsymbol{A}_{k}$ \texttt{for all} $k$;
	 $\text{SEE}_{out}=\text{SEE}(\tilde{\boldsymbol{p}},\tilde{\boldsymbol{\gamma}},\boldsymbol{C})$
		\UNTIL{$|\text{SEE}_{out}-\text{SEE}_{in}|<\epsilon$}
	\end{algorithmic}
\end{algorithm}

\begin{proposition}\label{Prop:SCA_Alt}
	Algorithm 3 monotonically increases the SEE value and converges in the value of the objective.
\end{proposition}

\begin{IEEEproof} 
Propositions~\ref{Prop:SCA_gamma1} and~\ref{Prop:SCA_p1} imply that Algorithm~\ref{Alg:SEE1} monotonically increases the SEE. Thus, convergence holds since the SEE has a finite maximizer.
\end{IEEEproof}

\textbf{Computational complexity:}  Neglecting the complexity of computing the receive filters $\boldsymbol{C}$, which are given in a closed-form expression, the complexity of Algorithm~\ref{Alg:SEE1} can be obtained recalling that pseudo-concave maximizations with $n$ variables can be restated as concave maximization with $n+1$ variables and thus their complexity is polynomial in $n+1$~\cite{ZapNow15}. Then, the optimization of RIS and power have complexity $\left(N+1\right)^\alpha$ and $\left(K+1\right)^\beta$,  respectively \footnote{The exponents $\alpha$ and $\beta$ are not known, but for generic convex problems they can be bounded between 1 and 4~\cite{BenTal2001ConvexOptimization}}, and thus the complexity of Algorithm~\ref{Alg:SEE1} is $\mathcal{C}_{1}=\mathcal{O}\left(I_{1}\left(I_{\gamma,1}\left(N+1\right)^\alpha + I_{p,1}\left(K+1\right)^\beta\right)\right)$ with  $I_{\gamma,1}$, $I_{p,1}$, $I_{1}$ the number of iteration for Algorithms~\ref{Alg:SCA_gamma1},~\ref{Alg:SCA_p1}, ~\ref{Alg:SEE1} to converge.

\section{Numerical Analysis}
\begin{figure}[!h]
	\centering
	\includegraphics[width=0.45\textwidth]{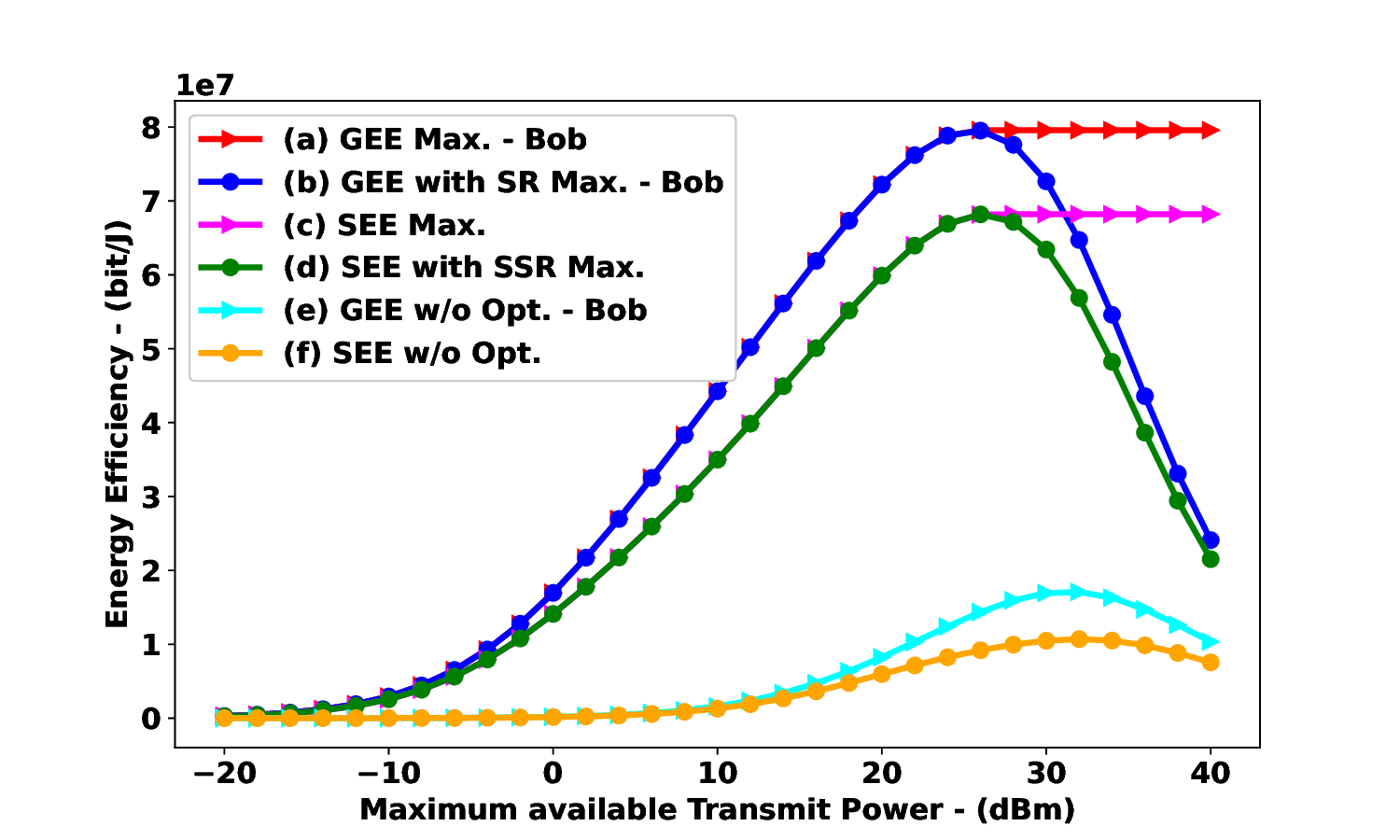}\caption{SEE versus $P_{t,max}$. $K=4$, $N_{B}=4$, $N_{E}=1$, $N=100$, $n_{h}=n_{g,E}=4, n_{g,B}=2$.} \label{fig:SEEvsP}
\end{figure}

Consider the setup detailed in Section~\ref{Sec:SysModel}. The parameters are set as follows:  $K=4$, $N_{B}=4$, $N_{E}=1$, $N=100$, $B=20\,\textrm{MHz}$, $P_{0}=30\,\textrm{dBm}$, $P^{(a)}_{0,RIS}=20\,\textrm{dBm}$, and $P^{(p)}_{c,n}=0\,\textrm{dBm}$. The noise spectral density is  $-174\,\textrm{dBm/Hz}$ with a noise figure of $5\,\textrm{dB}$. The mobile users are distributed within a $30\,\textrm{m}$ radius, maintaining a minimum distance of $R_{n}=20\,\textrm{m}$ from the RIS. Bob is positioned $20\,\textrm{m}$ away from the RIS, while Eve represents any potential eavesdropper within a $30\,\textrm{m}$ radius from Bob. The users exhibit height variations up to $2.5\,\textrm{m}$, whereas both the RIS and BS are elevated at $10\,\textrm{m}$. The power decay exponent of the users-RIS and  RIS-Eve channels 
is set to $n_{h}=n_{g,E}=4$, due to the worse propagation conditions caused by the mobility of the users and the fact that Eve might not be a legitimate user of the network and thus might be in an unfavorable position as far as signal reception is concerned. Instead, the RIS-Bob channel benefits from a reduced exponent $n_{g,B}=2$, motivated by the consideration that the legitimate receiver and the RIS are two fixed network nodes, whose position can be chosen in order to have a strong signal reception.
Fig.~\ref{fig:SEEvsP} shows the SEE versus $P_{t,max}$. Line (c) shows the SEE achieved by the proposed Algorithm~\ref{Alg:SEE1} for SEE maximization, while Line (a) shows the EE obtained assuming the ideal case in which the eavesdropper is not present. Lines (b) and (d) consider the same scenarios as Lines (a) and (c), respectively, with the difference that Line (d) shows the SEE obtained by the resource allocation that maximizes the secrecy sum rate (SSR), while line (b) shows the system EE obtained by the resource allocation that maximizes the system sum-rate assuming that no eavesdropper is present. Finally, Lines (f) and (e) consider the baseline scenario in which the phase values are randomly chosen in the interval $\left[0,\, 2\pi\right]$, along with uniform transmit powers.
Line (f) shows the SEE, while Line (e) shows the corresponding EE assuming that no eavesdropper is present. It is seen that, for all the schemes, the presence of an eavesdropper inevitably causes a slight decrease in the system's secrecy rate and EE, but the performance of the proposed Algorithm~\ref{Alg:SEE1} significantly outperforms the baseline scheme.

\begin{figure}[!h]
	\centering
	\includegraphics[width=0.45\textwidth]{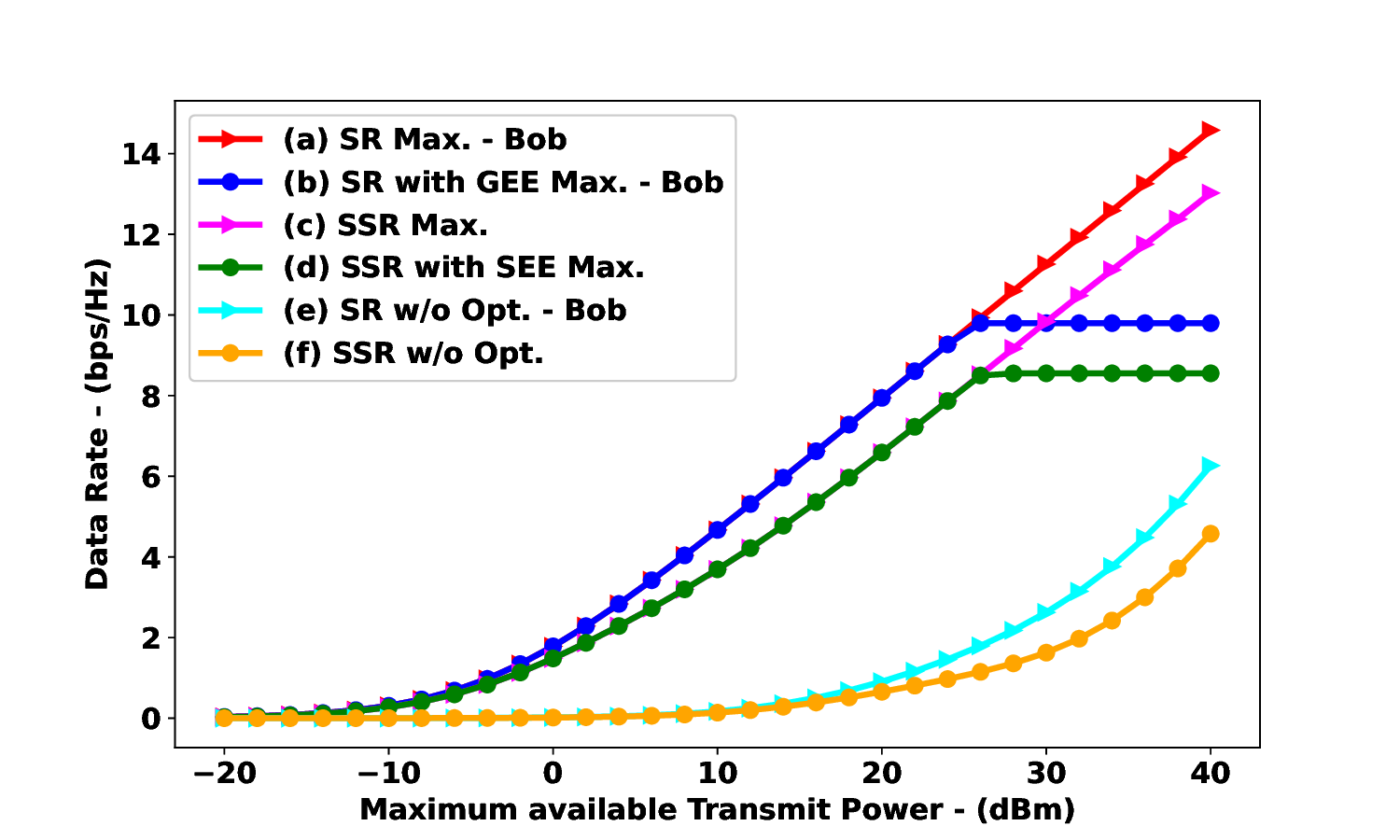}\caption{SSR versus $P_{t,max}$. $K=4$, $N_{B}=4$, $N_{E}=1$, $N=100$, $n_{h}=n_{g,E}=4, n_{g,B}=2$.} \label{fig:SSRvsP}
\end{figure}

A similar scenario is considered in Fig.~\ref{fig:SSRvsP}, in which the same resource allocation scenarios are shown. However, the metric that is reported is the system secrecy rate (SSR), for the schemes that consider the presence of the eavesdropper, and the system sum-rate, for the schemes that assume the case without eavesdropper. Similar considerations as for Fig.~\ref{fig:SEEvsP} can be made.

\begin{figure}[!h]
	\centering
	\includegraphics[width=0.45\textwidth]{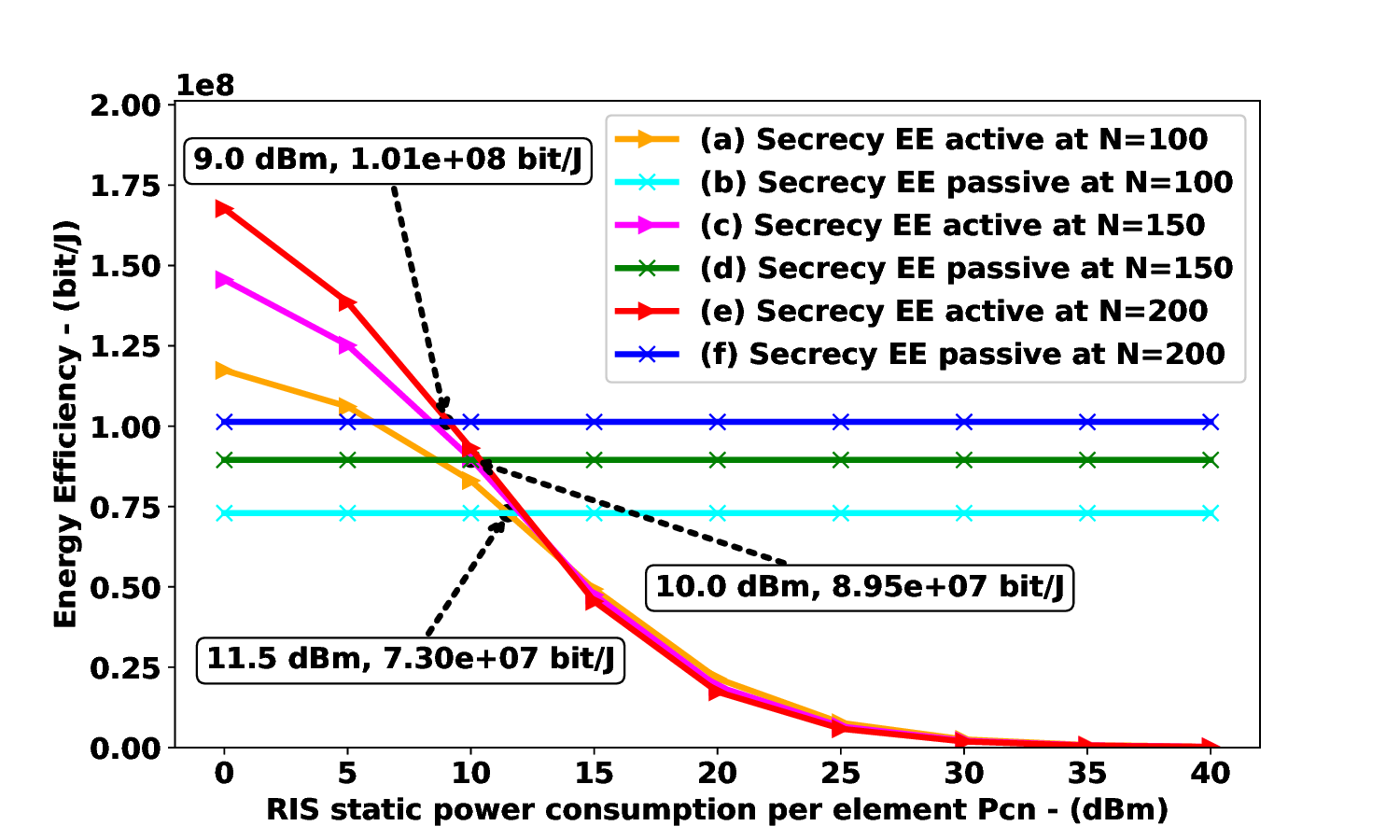}\caption{SEE for active and nearly-passive RIS versus the static power consumption. $K=4$, $N_{B}=4$, $N_{E}=1$, $N=(100,150,200)$, $P_{t,max}=30\,\textrm{dBm}$,  $P^{(a)}_{c,n}=10\,\textrm{dBm}$, $P^{(p)}_{c,n}=0\,\textrm{dBm}$, $P^{(a)}_{0,RIS}=20\,\textrm{dBm}$, and $P^{(p)}_{0,RIS}=10\,\textrm{dBm}$.} \label{fig:EEvsPcn}
\end{figure}

Fig.~\ref{fig:EEvsPcn} compares the SEE obtained by Algorithm~\ref{Alg:SEE1} with an active and a nearly-passive RIS. Specifically, it is assumed that the nearly-passive RIS has a per-element static power consumption of $P^{(p)}_{c,n}=0\,\textrm{dBm}$, while the per-element power consumption of the active RIS is assumed to vary between  $P^{(a)}_{c,n}=0\,\textrm{dBm}$ to  $P^{(a)}_{c,n}=40\,\textrm{dBm}$. It is seen that the SEE obtained by the active RIS degrades as the per-element power consumption or the number of elements increases, and it is interesting to notice that the nearly-passive RIS becomes rapidly more energy-efficient than its active counterpart as the power consumption of the active RIS increases.

\vspace{-0.5cm}

\vspace{-0.5cm}
\section{Conclusion}
This work has proposed a provably convergent, low complexity algorithm for SEE maximization in the uplink of a wireless network. The analysis has shown that the use of an active RIS is not always preferable to a nearly-passive RIS, as far as the SEE is concerned. 
The case of perfect CSI for the RIS-Eve channel has been considered, which is motivated if the eavesdropper is another node of the network (e.g. the base station of another operator), which inevitably receives the message and could theoretically decode its content. However, if the eavesdropper is a hidden node, the assumption of perfect CSI should be relaxed and partial CSI for the channel between the RIS and Eve should be assumed. 


\bibliographystyle{IEEEtran}
\bibliography{FracProg.bib, references.bib}

\begin{thebibliography}{10}
\providecommand{\url}[1]{#1}
\csname url@samestyle\endcsname
\providecommand{\newblock}{\relax}
\providecommand{\bibinfo}[2]{#2}
\providecommand{\BIBentrySTDinterwordspacing}{\spaceskip=0pt\relax}
\providecommand{\BIBentryALTinterwordstretchfactor}{4}
\providecommand{\BIBentryALTinterwordspacing}{\spaceskip=\fontdimen2\font plus
\BIBentryALTinterwordstretchfactor\fontdimen3\font minus
  \fontdimen4\font\relax}
\providecommand{\BIBforeignlanguage}[2]{{%
\expandafter\ifx\csname l@#1\endcsname\relax
\typeout{** WARNING: IEEEtran.bst: No hyphenation pattern has been}%
\typeout{** loaded for the language `#1'. Using the pattern for}%
\typeout{** the default language instead.}%
\else
\language=\csname l@#1\endcsname
\fi
#2}}
\providecommand{\BIBdecl}{\relax}
\BIBdecl

\bibitem{DiRenzo2020}
M.~{Di Renzo} \emph{et~al.}, ``Analytical modeling of the path-loss for
  reconfigurable intelligent surfaces – anomalous mirror or scatterer ?'' in
  \emph{2020 IEEE 21st International Workshop on Signal Processing Advances in
  Wireless Communications (SPAWC)}, 2020, pp. 1--5.

\bibitem{SmartWireless}
M.~Di~Renzo \emph{et~al.}, ``Smart radio environments empowered by
  reconfigurable {AI} meta-surfaces: An idea whose time has come,''
  \emph{EURASIP Journal on Wireless Communications and Networking}, vol. 2019,
  no.~1, pp. 1--20, 2019.

\bibitem{RuiZhang_COMMAG}
Q.~Wu and R.~Zhang, ``Towards smart and reconfigurable environment: Intelligent
  reflecting surface aided wireless network,'' \emph{IEEE Communications
  Magazine}, vol.~58, no.~1, pp. 106--112, 2020.

\bibitem{Huang2019}
C.~Huang \emph{et~al.}, ``Indoor signal focusing with deep learning designed
  reconfigurable intelligent surfaces,'' in \emph{2019 IEEE 20th International
  Workshop on Signal Processing Advances in Wireless Communications (SPAWC)},
  2019, pp. 1--5.

\bibitem{Lopezperez2021survey}
D.~Lopez-Perez \emph{et~al.}, ``A survey on {5G} radio access network energy
  efficiency: Massive {MIMO}, lean carrier design, sleep modes, and machine
  learning,'' \emph{IEEE Communications Surveys and Tutorials}, vol.~24, no.~1,
  pp. 653--697, 2022.

\bibitem{Long2021}
R.~Long, Y.-C. Liang, Y.~Pei, and E.~G. Larsson, ``Active reconfigurable
  intelligent surface-aided wireless communications,'' \emph{IEEE Transactions
  on Wireless Communications}, vol.~20, no.~8, pp. 4962--4975, 2021.

\bibitem{Fotock2023}
R.~K. Fotock \emph{et~al.}, ``Energy efficiency optimization in {RIS}-aided
  wireless networks: Active versus nearly-passive {RIS} with global reflection
  constraints,'' \emph{IEEE Transactions on Communications}, vol.~72, no.~1,
  pp. 257--272, 2024.

\bibitem{ZapTSP13}
A.~Zappone, P.~Cao, and E.~A. Jorswieck, ``Energy efficiency optimization in
  relay-assisted {MIMO} systems with perfect and statistical {CSI},''
  \emph{IEEE Transactions on Signal Processing}, vol.~62, no.~2, pp. 443--457,
  2014.

\bibitem{Li2023}
X.~Li, Y.~Zheng, M.~Zeng, Y.~Liu, and O.~A. Dobre, ``Enhancing secrecy
  performance for {STAR-RIS NOMA} networks,'' \emph{IEEE Transactions on
  Vehicular Technology}, vol.~72, no.~2, pp. 2684--2688, 2023.

\bibitem{Pei2023}
Y.~Pei \emph{et~al.}, ``Secrecy outage probability analysis for downlink
  {RIS-NOMA} networks with {O}n-{O}ff control,'' \emph{IEEE Transactions on
  Vehicular Technology}, vol.~72, no.~9, pp. 11\,772--11\,786, 2023.

\bibitem{Trigui2021}
I.~Trigui, W.~Ajib, and W.-P. Zhu, ``Secrecy outage probability and average
  rate of {RIS}-aided communications using quantized phases,'' \emph{IEEE
  Communications Letters}, vol.~25, no.~6, pp. 1820--1824, 2021.

\bibitem{Xu2021}
P.~Xu, G.~Chen, G.~Pan, and M.~Di~Renzo, ``Ergodic secrecy rate of
  {RIS}-assisted communication systems in the presence of discrete phase shifts
  and multiple eavesdroppers,'' \emph{IEEE Wireless Communications Letters},
  vol.~10, no.~3, pp. 629--633, 2021.

\bibitem{Zhang2022}
Z.~Zhang \emph{et~al.}, ``On the secrecy design of {STAR-RIS} assisted uplink
  {NOMA} networks,'' \emph{IEEE Transactions on Wireless Communications},
  vol.~21, no.~12, pp. 11\,207--11\,221, 2022.

\bibitem{Wei2022}
L.~Wei, K.~Wang, C.~Pan, and M.~Elkashlan, ``Secrecy performance analysis of
  {RIS}-aided communication system with randomly flying eavesdroppers,''
  \emph{IEEE Wireless Communications Letters}, vol.~11, no.~10, pp. 2240--2244,
  2022.

\bibitem{Yang2020}
L.~Yang, J.~Yang, W.~Xie, M.~O. Hasna, T.~Tsiftsis, and M.~Di~Renzo, ``Secrecy
  performance analysis of {RIS}-aided wireless communication systems,''
  \emph{IEEE Transactions on Vehicular Technology}, vol.~69, no.~10, pp.
  12\,296--12\,300, 2020.

\bibitem{Zhao2023}
M.-M. Zhao, K.~Xu, Y.~Cai, Y.~Niu, and L.~Hanzo, ``Secrecy rate maximization of
  {RIS}-assisted {SWIPT} systems: A {T}wo-{T}imescale beamforming design
  approach,'' \emph{IEEE Transactions on Wireless Communications}, vol.~22,
  no.~7, pp. 4489--4504, 2023.

\bibitem{Hoang2023}
T.~M. Hoang, C.~Xu, A.~Vahid, H.~D. Tuan, T.~Q. Duong, and L.~Hanzo,
  ``Secrecy-rate optimization of double {RIS}-aided space–ground networks,''
  \emph{IEEE Internet of Things Journal}, vol.~10, no.~15, pp.
  13\,221--13\,234, 2023.

\bibitem{Yizhi2023}
Y.~Li, Y.~Zou, J.~Zhu, B.~Ning, L.~Zhai, H.~Hui, Y.~Lou, and C.~Qin, ``Sum
  secrecy rate maximization for active {RIS}-assisted uplink {SIMO-NOMA}
  networks,'' \emph{IEEE Communications Letters}, pp. 1--1, 2023.

\bibitem{Cai2023}
X.~Cai, C.~Yuen, C.~Huang, W.~Xu, and L.~Wang, ``Toward chaotic secure
  communications: An {RIS} enabled {M-Ary} differential chaos shift keying
  system with block interleaving,'' \emph{IEEE Transactions on Communications},
  vol.~71, no.~6, pp. 3541--3558, 2023.

\bibitem{Yichi2023}
Y.~Zhang, Y.~Lu, R.~Zhang, B.~Ai, and D.~Niyato, ``Deep reinforcement learning
  for secrecy energy efficiency maximization in {RIS}-assisted networks,''
  \emph{IEEE Transactions on Vehicular Technology}, vol.~72, no.~9, pp.
  12\,413--12\,418, 2023.

\bibitem{Yang2023}
Y.~Lu, ``Secrecy energy efficiency in {RIS}-assisted networks,'' \emph{IEEE
  Transactions on Vehicular Technology}, vol.~72, no.~9, pp. 12\,419--12\,424,
  2023.

\bibitem{Hao23}
W.~Hao \emph{et~al.}, ``Max-{M}in security energy efficiency optimization for
  {RIS}-aided cell-free networks,'' in \emph{ICC 2023 - IEEE International
  Conference on Communications}, 2023, pp. 5358--5363.

\bibitem{Renzo2022}
M.~Di~Renzo, F.~H. Danufane, and S.~Tretyakov, ``Communication models for
  reconfigurable intelligent surfaces: From surface electromagnetics to
  wireless networks optimization,'' \emph{Proceedings of the IEEE}, vol. 110,
  no.~9, pp. 1164--1209, 2022.

\bibitem{Robert2023}
R.~K. Fotock, A.~Zappone, and M.~Di~Renzo, ``Energy efficiency in {RIS}-aided
  wireless networks: Active or passive {RIS}?'' in \emph{ICC 2023 - IEEE
  International Conference on Communications}, 2023, pp. 2704--2709.

\bibitem{ZapNow15}
A.~Zappone and E.~A. Jorswieck, ``Energy efficiency in wireless networks via
  fractional programming theory,'' \emph{Found. and Trends{\textregistered} in
  Commun. and Inf. Theory}, vol.~11, no. 3-4, pp. 185--396, 2015.

\bibitem{Mohamed2023}
M.~Elwekeil \emph{et~al.}, ``Power control in cell-free massive {MIMO} networks
  for {UAVs URLLC} under the finite blocklength regime,'' \emph{IEEE
  Transactions on Communications}, vol.~71, no.~2, pp. 1126--1140, 2023.

\bibitem{BenTal2001ConvexOptimization}
A.~Ben-Tal and A.~Nemirovski, \emph{Lectures on Modern Convex Optimization}, 01
  2012.

\end{thebibliography}

\end{document}